\documentclass{aastex}

\shorttitle{Dark-Matter-Dominated Galaxy Rotation Curves}
\shortauthors{Iliev and Shapiro}
\begin{document}
\title{On the Origin of the Rotation Curves of Dark-Matter-Dominated Galaxies} 
\author{Ilian T. Iliev
}
\affil{Department of Physics, University of Texas, Austin, 78712}
\email{iliev@astro.as.utexas.edu}
\and 
\author{Paul R. Shapiro}
\affil{Department of Astronomy, University of Texas, Austin, 78712}
\email{shapiro@astro.as.utexas.edu}
\slugcomment{Submitted to ApJ Letters}
\begin{abstract} 
Rotation curves of dark-matter-dominated galaxies measure the mass 
profiles of galactic halos and thereby test theories of 
their cosmological origin. While attention has focused lately 
on the possible discrepancy at small
galactocentric radii between observed rotation curves and
 the singular density profiles predicted by 
N-body simulations of the Cold Dark Matter (CDM) model, the observed 
rotation curves nevertheless contain valuable additional information with which to
test the theory and constrain the fundamental cosmological
parameters, despite this uncertainty at small radii.
 An analytical model we derived elsewhere 
for the postcollapse equilibrium of cosmological halos as
truncated, nonsingular, isothermal spheres (TIS) reproduces
many of the average properties of halos in CDM simulations to
good accuracy, including the density profiles outside the central region.
The circular velocity profile of this TIS model is, moreover,
 in excellent agreement with 
the observed ones and yields the mass and formation epoch of an 
observed halo from the  parameters of its rotation curve. This 
allows us to predict correlations amongst rotation curve parameters, 
such as the maximum velocity and the radius at which it 
occurs, for different mass halos forming at different epochs
in the CDM model. As an example, we derive the observed 
$v_{\rm max}$-$r_{\rm max}$ relation analytically, with preference 
for the flat $\Lambda$CDM model. 
\end{abstract}


\section{Introduction}
The rotation curves of dark-matter-dominated galaxies
probe the mass profile of their galactic halos. 
The observed rotation curves of dwarf and 
low-surface-brightness (LSB) disk galaxies, therefore,
offer a relatively direct test 
of theories of cosmological halo formation, free of the
complicating effects of the dynamical coupling of 
dissipationless dark matter and dissipative baryonic matter 
which affect the mass profiles of baryon-dominated galaxies. 
Recently, attention has focused on the apparent conflict
between the singular halo density profiles predicted by CDM
N-body simulations and the observed rotation curves,
which favor a flat-density core (cf. Moore et al. 1999). 
This has led to intense scrutiny of both the 
observations and the CDM model, primarily
in one of three directions: improvement in the numerical resolving 
power of the CDM N-body simulations to determine better
the logarithmic
slope of the predicted density profiles at small radii
(e.g. Moore et al. 1999), 
ideas for modifying the microscopic properties of the
dark matter so as to retain the more successful aspects of the
CDM model while flattening the halo density profiles at small 
radii (e.g. Dav\'e et al. 2000, and references therein), and suggestions 
that the rotation curve data lack sufficient spatial resolution 
near the center to distinguish unambiguously between a density profile 
with a flat-density core and the singular profiles predicted 
by CDM N-body simulations (e.g. van den Bosch \& Swaters 2000).
Despite these uncertainties, a meaningful comparison between the data 
and predictions for the {\it global} properties of the rotation curves 
of dark-matter-dominated galaxies is possible which serves to 
test the CDM model and discriminate amongst different background 
cosmologies. To demonstrate this, we show how the mass and formation 
epoch of a halo can be extracted from its observed rotation curve, by 
application of an analytical model for halo formation which reproduces the
empirical description of these rotation curves extremely well.

\cite{B95} showed that the observed rotation curves of 
several dark-matter-dominated dwarf galaxies are consistent with a 
common density profile with a flat-density core, according to 
\begin{equation}
\label{rho_B}
\rho(r)=\frac{\rho_{\rm 0,B}}{(r/r_{\rm 0,B}+1)(r^2/r_{\rm 0,B}^2+1)}.
\end{equation}
\cite{KKBP} found that the rotation 
curves for a larger sample which included both dwarf and LSB galaxies 
are well fit by a similar universal halo profile given by
\begin{equation}
\label{general_profile}
\displaystyle{\rho(r)
	=\frac{\rho_S}{({r}/{r_S})^\gamma
		(1+({r}/{r_S})^\alpha)^{(\beta-\gamma)/\alpha}}},
\end{equation}
with $(\alpha,\beta,\gamma)=(2,3,0.2)$, which is only slightly 
steeper than the 
Burkert profile in the core but shares the slope at large radii, where
$\rho\propto r^{-3}$. The circular velocity profiles 
[i.e. $v(r)\equiv(GM[\leq r]/r)^{1/2}$] for equations (\ref{rho_B}) and 
(\ref{general_profile})
are virtually indistinguishable at all radii,
even near the center, where $v\propto r^{1-\gamma/2}$, since 
$1-\gamma/2=1$ or ${0.9}$, respectively. 

This contrasts with the halo profiles found by N-body 
simulations of the standard CDM model. The universal fitting formula 
for simulated halos reported 
by Navarro, Frenk, \& White (1997; NFW) is equation~(\ref{general_profile})
with $(\alpha,\beta,\gamma)=(1,3,1)$, while 
\citet{MQGSL} find  
$(\alpha,\beta,\gamma)=(1.5,3,1.5)$.
However, the circular velocity profiles for these halos 
differ significantly from those implied by equations (\ref{rho_B}) and 
(\ref{general_profile}) only at very small radii where
 the uncertainties in the shape of
the observed rotation curves currently makes discrimination difficult.
In the meantime, 
equation~(\ref{rho_B}) generally fits the data 
better than does the NFW profile even when the latter is also
an acceptable fit.

The Burkert profile, then, continues to serve as a useful empirical 
description
of the universal mass profiles of dark-matter-dominated galactic halos.
We show here that the truncated, nonsingular isothermal sphere model
we derived elsewhere (Shapiro, Iliev \& Raga 1998, Paper I; 
Iliev \& Shapiro 2000, Paper II) has a density profile for which 
the circular velocity profile is
 essentially indistinguishable from that of the Burkert profile
and, as such, provides a theoretical motivation for the latter. The TIS model
goes well beyond the prediction of density profile and rotation curve, however,
to provide the size, mass, velocity dispersion and collapse epoch of the halo,
as well. This makes possible 
the further interpretation of observations of dark-matter-dominated galaxies 
for comparison with the predictions of various cosmological models. At the
same time, the TIS model can be shown to reproduce many of the average 
properties of the halos found in simulations of the standard CDM model, outside
of the innermost region where the TIS halo has a flat-density core, unlike the
CDM halos. As a result, a comparison of the 
analytical TIS predictions with observed properties of galaxies also offers
 insight into the standard CDM model. Finally, if  
suggestions like the self-interacting dark matter proposal of 
\cite{SS} are correct, that CDM might be more ``collisional'' 
as a way to eliminate the central cusp of standard CDM halos, then our 
TIS solution will also apply to these models, to the 
extent that the halo relaxation process makes
the final equilibrium approximately isothermal. 

In \S~\ref{tis}, we briefly summarize the relevant properties of the
TIS model and present a simple analytical formula for it with which 
to fit an observed rotation curve. In \S~\ref{applic_rot_curves} we
demonstrate the excellent agreement between the TIS model rotation curve 
and that implied by equation~(\ref{rho_B}) and show how this allows us 
to deduce the total mass and collapse epoch of a given halo directly
from the parameters of its observed rotation curve. In 
\S~\ref{vm_rm_sect}, we describe how the dependence of the average 
formation epoch of a halo on its mass in the CDM model results in 
statistical correlations amongst the parameters of the observed rotation 
curves. As an example, we use this approach to derive analytically 
 the known correlation 
between the maximum velocity of each rotation curve and the radius at which
it occurs.

\section{The Truncated Isothermal Sphere Model}
\label{tis}
The TIS model is a 
 particular solution of the Lane-Emden equation
(suitably modified when $\Lambda\neq0$) which results from the collapse and 
virialization of a top-hat density perturbation
(c.f. Paper I for Einstein-de~Sitter [EdS] universe and
Paper II for $\Omega_0<1$ and $\lambda_0=0$ or $1-\Omega_0$). The size $r_t$
and velocity dispersion $\sigma_V$
are unique functions of the mass and redshift of formation of the 
object for a given background universe. 
While the Lane-Emden equation 
requires a straightforward numerical solution, Paper I provides a convenient 
analytical fitting formula,
\begin{equation}
\label{analyt_fit_rho}
\displaystyle{\rho(r)=\rho_0\left[\frac{A}{a^2+\zeta^2}
			-\frac{B}{b^2+\zeta^2}\right],}
\end{equation}
where $\zeta\equiv r/r_0$ and $(A,a^2,B,b^2)=(21.38,9.08,19.81,14.62)$, 
accurate to within 3\% 
over the full range $0\leq \zeta\leq \zeta_t\approx30$ for both the EdS case
and most low-density models of interest (i.e. $\Omega_0\geq0.3$).
[Note: Our definition of the core radius 
is $r_{\rm 0,TIS}\equiv r_{\rm King}/3$, where $r_{\rm King}$ is the
``King radius'' defined in Binney \& Tremaine (1987), p. 228.]
Equation~(\ref{analyt_fit_rho}) can be integrated to yield an analytical
fitting formula for the TIS rotation curve, as well, given by
\begin{equation}
\label{fit_v}
\displaystyle{\frac{v(r)}{\sigma_V}
=\left\{A-B
+\frac{1}{\zeta}\left[bB\tan^{-1}\left(\frac{\zeta}{b}\right)\right.\right.}\nonumber\\
\displaystyle{\left.\left.-aA\tan^{-1}\left(\frac{\zeta}{a}\right)\right]\right\}^{1/2}},
\end{equation}
where $\sigma_V=(4\pi G\rho_0r_0^2)^{1/2}$.
This fit has a fractional error less than 1\% over the full range of radii
$0\leq r\leq r_t$ for all matter-dominated background models. 
With a nonzero cosmological constant, the circular velocity becomes 
$v(r)=(GM[\leq r]/r)^{1/2}[1-2\rho_\lambda/\rho(r)]^{1/2}$, where $\rho_\lambda$
is the constant vacuum energy density associated with the cosmological constant.
For cases of current interest for a flat universe with $\Lambda\neq0$ 
(i.e. $\Omega_0\geq0.3$), equation~(\ref{fit_v}) can still be
used, however, since it departs from the exact solution
only slightly in the outer halo (i.e. at 
$r\sim r_t$) for halos collapsing even as late as $z=0$; the fractional 
error is less than 6\% at all radii and less than 1\% for $r\leq(2/3)r_t$.

The TIS model
quantitatively reproduces the average structural properties of
halos found in CDM simulations to good accuracy, suggesting that it is a 
useful analytical approximation for halos which form from more realistic
initial conditions. An exception to this agreement is the very inner 
profile, where the TIS has a uniform-density core instead of a 
central cusp. Our TIS predictions agree to 
astonishingly high accuracy (i.e. to of order 1\%) with the cluster
mass-radius and radius-temperature relationships and integrated mass 
profiles derived from detailed CDM simulations of X-ray cluster formation
by Evrard, Metzler, and Navarro (1996). Apparently,
these simulation results are not sensitive to our disagreement in the 
core. A direct 
comparison of the TIS and NFW mass profiles reveals a very close agreement 
(fractional deviation of less than $\sim 10\%$)
at all radii outside of a few TIS core radii (i.e. about one King radius), 
for NFW concentration parameters $4\lesssim c_{\rm NFW}\lesssim7$.

\section{Application to Galaxy Rotation Curves} 
\label{applic_rot_curves}
The rotation curve for  
equation~(\ref{rho_B}) is given by
\begin{equation}
\label{v_B}
\displaystyle{\frac{v_{\rm B}(r)}{v_{\rm *,B}}}=
	\displaystyle{\left\{\frac{\ln\left[(\zeta_B+1)^2
	(\zeta_B^2+1)\right]
	-2\tan^{-1}(\zeta_B)}{\zeta_B}\right\}^{1/2}},
\end{equation} 
where $\zeta_B\equiv r/r_{\rm 0,B}$ and 
$v_{\rm *,B}\equiv(\pi G\rho_{\rm 0,B}r_{\rm 0,B}^2)^{1/2}$. 
To compare our TIS directly with equation~(\ref{rho_B}),
we solve for the ratios $\rho_{\rm 0,TIS}/\rho_{\rm 0,B}$ and 
$r_{\rm 0,TIS}/r_{\rm 0,B}$ which minimize the $\chi^2$ of the fit of 
equation~(\ref{v_B}) to equation~(\ref{fit_v}) over radii from
0 to $r_t$. The result in Figure~\ref{fit}, with 
${\rho_{\rm 0,TIS}}/{\rho_{\rm 0,B}}=0.7790$ and ${r_{\rm 0,TIS}}/{r_{\rm 0,B}}=0.3276$,
shows extremely good agreement, with a fractional deviation below 10\% 
from $0.01r_t$ to $r_t$ and below 4\% over the range 
 $r>0.03r_t\approx r_{\rm 0,TIS}$.
For the TIS, the maximum circular velocity and its location
are $v_{\rm max,TIS}=1.5867\sigma_{\rm V,TIS}$ at 
$r_{\rm max,TIS}=8.99r_{\rm 0,TIS}$ for all matter-dominated cosmologies. For
a flat universe with $\Lambda\neq0$, ($\Omega_0\geq0.3$), these numbers depend
weakly upon collapse redshift, but are reduced
by no more than 0.2\% and 1.7\%, respectively, for $z_{\rm coll}=0$, and by 
even less for earlier collapse.
For the Burkert profile, $v_{\rm max,B}=1.2143v_{\rm *,B}$ at 
$r_{\rm max,B} = 3.2446\,r_{\rm 0,B}$. Hence, our best fit finds that 
$v_{\rm max}$ and $r_{\rm max}$ for the TIS and Burkert profiles are extremely 
close, with ${v_{\rm max,TIS}}/{v_{\rm max,B}}=1.02$ and 
${r_{\rm max,TIS}}/{r_{\rm max,B}}=0.91$.

In short, our TIS model provides a solid, theoretical underpinning
for the empirical fitting formula of Burkert, and, by extension, a self-consistent
theoretical explanation of the observed galaxy rotation curves it was invented
to fit. In addition, the TIS model allows us to calculate the total mass $M_0$,
collapse epoch $z_{\rm coll}$, and other parameters of each observed
dark-matter-dominated halo from its rotation curve, as follows. 
Assuming $v_{\rm max}$ and $r_{\rm max}$ are provided by observation\footnote{
 In practice, observations are generally restricted to the inner parts of galaxy 
 rotation curves, often not extending to radii as large as $r_{\rm max}$. In that 
 case, $v_{\rm max}$ and $r_{\rm max}$ are inferred by fitting the observed 
 rotation curve to the TIS rotation curve at other radii.}, 
the mass of the galaxy can be shown to be
\begin{equation}
\label{M0}
\displaystyle{\frac{M_0}{h^{-1}M_{\rm \odot}}}=
   \displaystyle{6.329\times10^{10}\left(\frac{r_{\rm max}}{10\,h^{-1}\,{\rm kpc}}\right)
	\left(\frac{v_{\rm max}}{100\,{\rm km\,s^{-1}}}\right)^2}
\end{equation}
for any matter-dominated universe.
When $\Lambda\neq0$, $M_0\propto r_{\rm max}v_{\rm max}^2$, 
but the coefficient depends on $z_{\rm coll}$ and background cosmology 
(Paper II). For any flat universe ($\Lambda\neq0$) of current interest,
equation~(\ref{M0}) is still a very good approximation.
For $\Omega_0=1-\lambda_0\geq0.3$, 
equation~(\ref{M0}) underestimates the mass by less than 
6.3\%, 2.8\%, and 1.2\% for $z_{\rm coll}=0,0.5$, and 1, respectively. 

In terms of $r_{\rm max}$ and $v_{\rm max}$, $z_{\rm coll}$ is given to 
good accuracy by the implicit equation
\begin{equation}
\label{zcoll}
F(\Omega_0,\lambda_0,z_{\rm coll})=
   \displaystyle{2.284\left(\frac{v_{\rm max}/100\,{\rm km\,s^{-1}}}{
	{r_{\rm max}}/{10\,h^{-1}\,{\rm kpc}}}\right)^{2/3}}
\end{equation}
where 
$F\equiv\{[\Omega_0/\Omega(z_{\rm coll})][\Delta_{\rm c,SUS}/18\pi^2]\}^{1/3}
(1+z_{\rm coll})$, $\Omega(z)=[\Omega_0(1+z)^3]/[(1-\Omega_0-\lambda_0)(1+z)^2
+\Omega_0(1+z)^3+\lambda_0]$, and $\Delta_{\rm c,SUS}$ is the density [in units of 
$\rho_{\rm crit}(z_{\rm coll})$] after top-hat collapse and virialization
for the standard uniform sphere approximation (SUS)\footnote{This 
$\Delta_{\rm c,SUS}$ is well-approximated by $\Delta_{\rm c,SUS}=18\pi^2+c_1x-c_2x^2$,
where $x\equiv\Omega(z_{\rm coll})-1$, and $c_1=82\, (60)$ and $c_2=39\,(32)$ for the
flat (open) cases, $\Omega_0+\lambda_0=1$ ($\Omega_0<1,\lambda_0=0$), respectively 
\citep{BN}.}. For the EdS case, $F=(1+z_{\rm coll})$, while for 
open, matter-dominated and flat cases, $F\rightarrow\Omega_0^{-1/3}(1+z_{\rm coll})$ 
at early times [i.e. $x\rightarrow0$]. The coefficient on the r.h.s. of 
equation~(\ref{zcoll}) is correct for any matter-dominated cosmology and a very good 
approximation for flat ($\Lambda\neq0$) cases of current interest. For
$\Omega_0\geq0.3$, it underestimates $(1+z_{\rm coll})$ by less than 2.5\% for 
$z_{\rm coll}\geq0$.

\section{Statistical Correlations Amongst the Properties of Dwarf and
LSB Galaxies: The $v_{\rm max}$--$r_{\rm max}$ Relation}
\label{vm_rm_sect}
The halos in our TIS model are fully described for a given background universe
by their total mass $M_0$ and collapse epoch $z_{\rm coll}$. In hierarchical 
models of structure formation like CDM, these are not statistically 
independent 
parameters, however. Smaller mass halos on average collapse earlier and are 
denser than larger mass halos. In terms of galaxy rotation-curve parameters, 
this dependence should be observable, for example, as a correlation between 
$v_{\rm max}$ and $r_{\rm max}$. \citet{MB} used the \cite{B95} fits to
rotation curves of dwarf galaxies to report such an observed
correlation, expressed as follows:
\begin{equation}
\label{scalingB2}
\displaystyle{v_{\rm max,B} 
  = 9.81\left({r_{\rm max,B}}/{1 \,{\rm kpc}}\right)^{2/3}\,{\rm km\,s^{-1}}}.
\end{equation}
\cite{KKBP} also found a correlation using fits to observed
rotation curves. Their results for a sample of dwarf and LSB galaxies 
 are shown in Figure~\ref{vm-rm}, along with that in equation~(\ref{scalingB2}). 
 \cite{KKBP} further showed that the results of their CDM N-body simulations
 agreed with these data points.\footnote{This result by \cite{KKBP}
 is not sensitive to the question of whether their simulations adequately 
 resolved the halo density profiles at very small radii.}
This suggests that if we can relate
the mass and collapse epoch of our TIS model halos in a statistical way within 
the context of the CDM model, a comparison of our predicted $v_{\rm max}$-$r_{\rm max}$ 
relation with this observed one will further check the relevance of
our TIS model to halo formation from realistic initial conditions and, at the 
same time, give a theoretical explanation for both the data and the simulation results.
 
To predict the $v_{\rm max}$--$r_{\rm max}$ correlation for a CDM universe
using our TIS model, we apply the well-known 
Press-Schechter (PS) approximation to derive $z_{\rm coll}(M_0)$ 
- the typical collapse epoch for a halo with a given mass.
Halos of mass $M$ which collapse when $\sigma(M)=\delta_{\rm crit}/\nu$ 
are referred to as ``$\nu$-$\sigma$'' fluctuations, where $\sigma(M)$ is
the standard deviation of the density fluctuations at $z_{\rm coll}$ according to 
linear perturbation theory, after the density field is filtered on the scale 
$M$, and $\delta_{\rm crit}$ is the amplitude of a top-hat perturbation according 
to linear theory at the epoch $z_{\rm coll}$ at which the exact solution predicts
infinite collapse. The typical collapse epoch for halos of a given mass is that
for which $\nu=1$, the  1-$\sigma$ fluctuations.

Our results for 1-$\sigma$ fluctuations are shown in Figure~\ref{vm-rm} 
(upper panel), for different background cosmologies.
The flat, untilted and the open, slightly-tilted 
($n_p=1.14$) models are in reasonable agreement with the observed 
$v_{\rm max}-r_{\rm max}$ relation, while the untilted
and strongly tilted ($n_p=1.3$) open models are not. The empirical Burkert 
scaling relation is closely approximated by the currently-favored 
$\Lambda$CDM model, less well by other models.  

In practice, observed galaxies should exhibit a statistical spread of halo 
properties in accord with the expectations of the Gaussian statistics of 
the density fluctuations which formed them. Since observed galaxies will 
not all be ``typical'', the scatter of the data points 
in Figure~\ref{vm-rm} is natural. To probe this in our model,
we calculate the masses and collapse redshifts for halos formed 
by $\nu$-$\sigma$ fluctuations for different values of $\nu$. 
As shown in Figure~\ref{vm-rm} (lower panel) for $\Lambda$CDM, the 
Burkert scaling relation is closest to the TIS model prediction for 1-$\sigma$
fluctuations, while all the observed galaxy data points except one 
correspond to $0.7\lesssim\nu\lesssim1.5$. Hence, on average, the galaxies 
which constitute
the observed $v_{\rm max}$-$r_{\rm max}$ correlation correspond to halos which formed 
at close to the typical collapse time expected theoretically for objects 
of that mass. This means that our TIS model is a self-consistent explanation 
for the observed $v_{\rm max}$-$r_{\rm max}$ correlation.

This success of the TIS model in explaining the observed $v_{\rm max}$-$r_{\rm max}$ 
relation and, by extension, the CDM simulation results of \cite{KKBP} which follow
it can be understood by a completely analytical argument, as follows. We approximate
the density fluctuation power spectrum as a power-law in wavenumber $k$, 
$P(k)\propto k^n$. If we define a mass $M$ which corresponds to $k$ according to 
$M\propto k^{-3}$, then $\sigma(M)\propto M^{-(3+n)/6}$ if we set 
$n=n_{\rm eff}\equiv-3(2y_{\rm eff}+1)$, where 
$y_{\rm eff}\equiv(d\ln \sigma/d\ln M)_{\rm exact}$ at the relevant
mass scale $M$. Our results for the
$\Lambda$CDM case indicate that the galaxies which
make up the $v_{\rm max}-r_{\rm max}$ data points in Figure~\ref{vm-rm} 
collapsed at redshifts $1\lesssim z_{\rm coll}\lesssim6$
with masses in the range 
$8\times10^9\lesssim M_0/(M_\odot h^{-1})\lesssim 3\times10^{11}$.
Hence, the precollapse fluctuation growth rate is approximately 
EdS, and we can let $\Omega(z_{\rm coll})=1$ in 
equations~(\ref{M0}) and (\ref{zcoll}). In that case, 
$(1+ z_{\rm coll})\propto\sigma(M)\propto M^{-(3+n)/6}$,
$r_{\rm max}\propto M^{(5+n)/6}\Omega_0^{-1/3}$ and 
$v_{\rm max}\propto M^{(1-n)/12}\Omega_0^{1/6}$, which combine to yield  
\begin{equation}
\label{vm_rm_scaling_analyt}
\displaystyle{v_{\rm max}=
	v_{\rm max,*}\left({r_{\rm max}}/{r_{\rm max,*}}\right)^{(1-n)/[2(5+n)]}},
\end{equation}
where $v_{\rm max,*}$ and $r_{\rm max,*}$ are for a 1-$\sigma$ 
fluctuation of fiducial mass $M_*$, as given by
equations~(\ref{M0}) and (\ref{zcoll}) with 
$(1+z_{\rm coll})=(1+z_{\rm rec})\sigma(M_*,z_{\rm rec})/\delta_{\rm crit}$,
where $\delta_{\rm crit}=1.6865$ and $\sigma(M_*,z_{\rm rec})$ is the value of
$\sigma(M_*)$ evaluated at the epoch of recombination $z_{\rm rec}$ [i.e. 
early enough that $(1+z)\sigma$ is independent of $z$].
Over the relevant mass range $M=10^{10\pm1}{ h^{-1}M_\odot}$, 
$n_{\rm eff}\approx-2.4\pm0.1$ 
for our $\Lambda$CDM case. For $M_*=10^{10}h^{-1}M_\odot$, our 
COBE-normalized, flat $\Lambda$CDM case ($\Omega_0=0.3,h=0.65$)
yields $(1+z_{\rm rec})\sigma(M_*,z_{\rm rec})=5.563$,
so $(1+z_{\rm coll})=3.30$, $v_{\rm max,*}=53.2\,{\rm km\,s^{-1}}$ and 
$r_{\rm max,*}=5.59\,h^{-1}{\rm kpc}$. With these values and 
$n_{\rm eff}=-2.4$, equation~(\ref{vm_rm_scaling_analyt}) yields the
TIS model analytical prediction 
$v_{\rm max}=(13.0\, {\rm km\,s^{-1}})(r_{\rm max}/1\,{\rm kpc})^{0.65}$, 
remarkably close to the empirical relation in equation~(\ref{scalingB2}).


\acknowledgments
We thank Hugo Martel for valuable discussion, grants NASA ATP NAG5-7363 
and NAG5-7821, NSF ASC-9504046, and Texas 
Advanced Research Program 3658-0624-1999, and an
NSF International Research Fellow Award INT-0003682 to ITI.

\figcaption[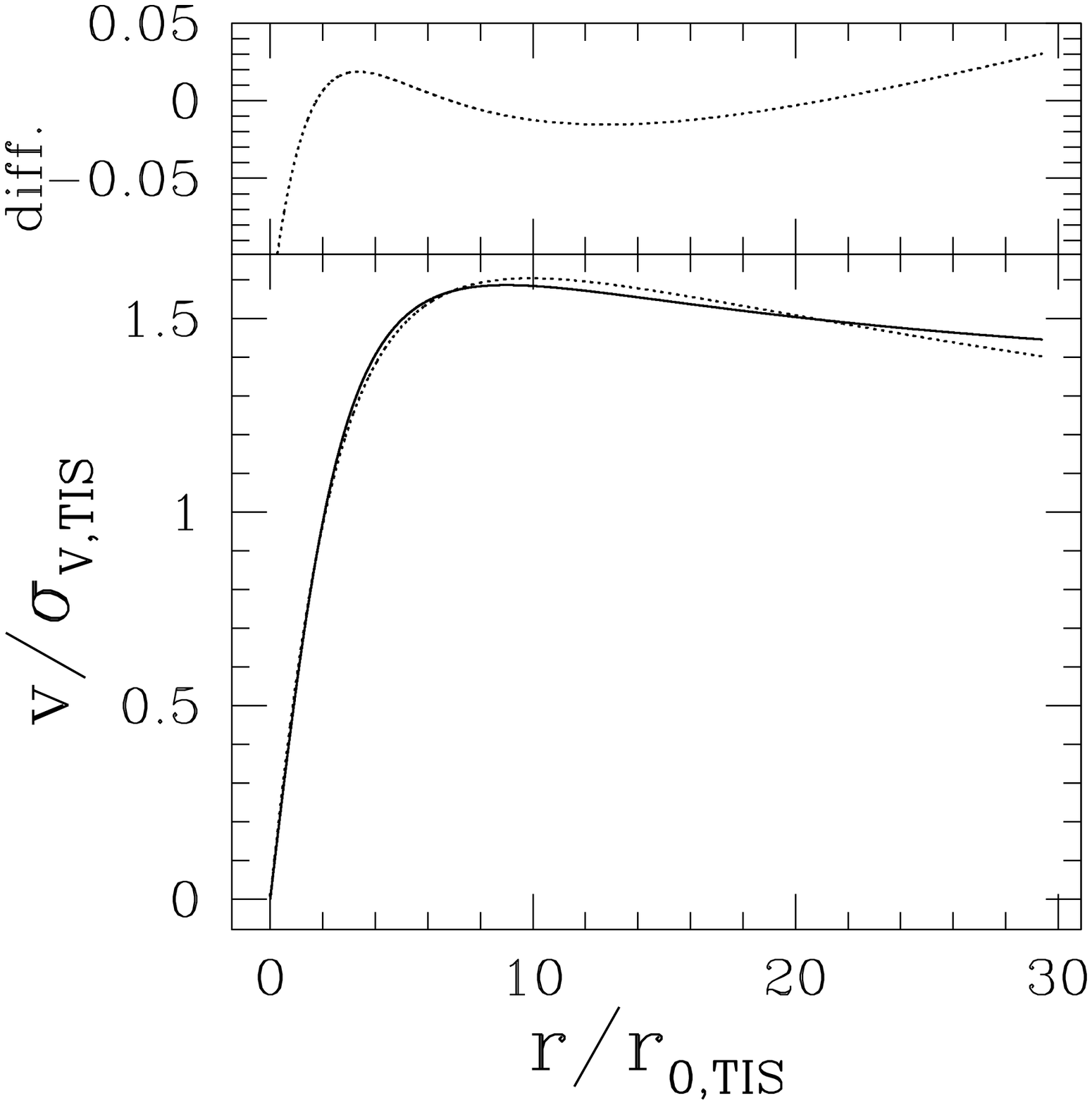]{Circular velocity profiles $v(r)$ (in units of 
$\sigma_{V,TIS}$) versus radius $r$ (in units of $r_{0,TIS}$)
for the TIS model halo (matter-dominated universe)(solid line) and for the
best-fit Burkert (1995) profile (dotted line), fit over the entire range 
$0\leq r\leq r_t$ (i.e. $0\leq\zeta\leq\zeta_t\approx 30$).
Top panel shows the fractional deviation $(v_{\rm TIS}-v_B)/v_{\rm TIS}$.
\label{fit}}
\figcaption[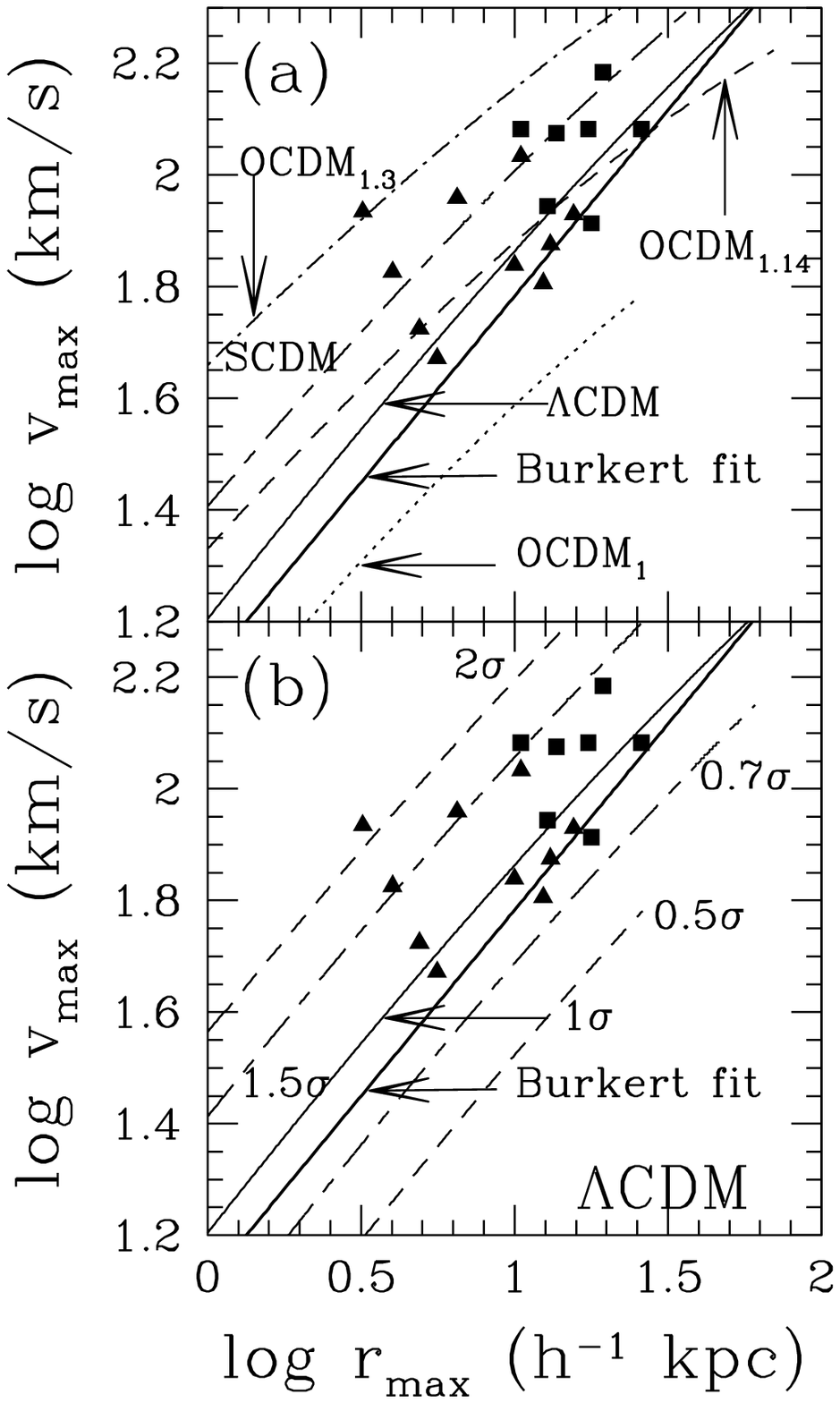]{(a) (upper panel)
$v_{\rm max}$--$r_{\rm max}$ correlation.
Observed dwarf galaxies (triangles) and LSB galaxies (squares) from 
Kravtsov et al. (1998). Curve labelled ``Burkert fit'' is 
equation~(\ref{scalingB2}) based on a fit by \cite{B95} to data.
 TIS+PS predictions are plotted for different CDM cosmologies, as labelled:
SCDM ($\Omega_0=1$, cluster-normalized: $\sigma_{8{\rm h^{-1}}}=0.5$), OCDM  
($\Omega_0=0.3$, $\lambda_0=0$, COBE-normalized, both untilted 
[primordial power spectrum index $n_p=1$] and tilted [$n_p=1.14$ and 
$1.3$] (subscripts indicate 
$n_p$-value); and $\Lambda$CDM ($\Omega_0=1-\lambda_0=0.3$, 
COBE-normalized). In the untilted $\Lambda$CDM and strongly-tilted 
($n_p=1.3$) open cases, COBE- and cluster-normalizations are the same. 
All models assume $h=0.65$.
(b) (lower panel)
Same as (a), except TIS+PS curves are for the $\Lambda$CDM 
model only, for fluctuations of different initial amplitudes for each mass
scale, as labelled by the ``$\nu$-$\sigma$'' values: 
1-$\sigma$ (i.e. ``typical'' fluctuations), 
0.5-$\sigma$,0.7-$\sigma$,1.5-$\sigma$, and 2-$\sigma$. \label{vm-rm}}

\begin{figure}
	\centering 
    \includegraphics[width=\textwidth]{fig1.eps}
\end{figure}

\begin{figure}
	\centering
    \includegraphics[width=1.35\textwidth]{fig2.eps}
\end{figure}

\end{document}